\documentclass[useAMS,usenatbib]{mnras}

\usepackage{bm}
\usepackage{natbib}
\usepackage{graphicx}
\usepackage{amssymb, amsmath}

\title[Simplified reaction network]{Crust of accreting neutron stars within simplified reaction network}
\author[N. N. Shchechilin, A. I. Chugunov]
{N. N. Shchechilin
\thanks{nicknicklas@mail.ru} and A. I. Chugunov\\
$^1$Ioffe Institute, Politekhnicheskaya 26, 194021 Saint Petersburg, Russia}

\begin{document}
	
	\date{Accepted 2019 xxxx. Received 2019 xxxx;
		in original form 2019 xxxx}
	
	\pagerange{\pageref{firstpage}--\pageref{lastpage}}
	\pubyear{2019}
	
	\maketitle
	
	\label{firstpage}

\begin{abstract}
Transiently accreting neutron stars in low mass X-ray binaries are generally believed to be heated up by nuclear reactions in accreted matter during hydrostatic compression. 
Detailed modeling of these reactions is required for the correct interpretation of observations.
In this paper, we construct a simplified reaction network, which can be easily implemented and  depends mainly on atomic mass tables as nuclear physics input.  We show that it reproduces results of the detailed network by Lau et al.\ (2018) very well, if one applies the same mass model.
However, the composition and the heating power are shown to be sensitive to the mass table used and treatment of mass tables boundary, if one applies several of them in one simulation. In  particular, the impurity parameter $Q_\mathrm{imp}$ at density $\rho=2\times 10^{12}$~g\,cm$^{-3}$ can differ for a factor of few, and
even increase with density increase.
The profile of integrated  heat realize  shown to be well confined between results by Fantina et al.\ (2018) and Lau et al. (2018).
\end{abstract}


\section{Introduction}

Many neutron stars are located in low mass X-ray binaries, where matter is transferred from Roche-lobe-filling companion star to the neutron star. On course of the accretion, neutron star can be detected as a powerful source of X-rays (luminosity $L\sim 10^{36}$~erg\,s$^{-1}$) and this emission is associated with the release of the gravitational energy of the accreted material. In some cases, the accretion process is not stable and sometime stops down, resulting in a decrease of X-ray luminosity for orders of magnitude. In these quiescent periods the emission, detected by X-ray telescopes, contains a component, which is associated with thermal emission from the surface  (e.g., \citealt*{wdp17}) directly or which, at least, provides an opportunity to derive an upper limits to the thermal component (e.g., \citealt{heinke_et_al_09}). 

It is generally believed,  that the heating for this emission is provided by nuclear reactions in the crust of the accreting neutron star, stimulated by compression of accreted material
(e.g.,\ \citealt*{bbr98}).
These reactions not only provide the necessary heating, but also determine the equation of state and composition. Both are crucial to model
the thermal emission, constraining the properties of neutron star's crust and core (e.g., \citealt{mdkse18}; \citealt*{pcc19}).

The temperature in the crust of accreting neutron star is not high enough, and reaction rate is therefore too small for a wide set of nuclear reactions, which are required to drive nuclei to nuclear statistical equilibrium. As a result, the composition evolves under the action of limited number of nuclear reactions and does not arrive to the equilibrium. Thus, special analysis is demanded. 

First calculations were performed by \cite{bkc74,Sato79}, and the most widely used model was suggested by  \cite{HZ90,HZ90b,HZ03,HZ08}. The latter was recently updated within more realistic nuclear physics model based on Hartree-Fock-Bogoliubov (HFB)  approach and energy-density functionals of generalized Skyrme type by \cite{Fantina_ea18}. However, the above models assume one component composition of the crust at each pressure. Importance of multicomponent effects was shown by \cite{Gupta_ea07}, and in their recent work \cite{lau_ea18} constructed the reaction network model for the neutron star crust up to densities $\sim 2\times10^{12}$~g\,cm$^{-3}$. Still, implementation of the multicomponent reaction network requires extensive computations and detailed knowledge of the reaction rates, which should be calculated consistently with the nuclear physics model. Thus it is rather difficult to analyze the effects of nuclear physics uncertainties within the full reaction network.

To overcome this difficulty we suggest the simplified reaction network model, which appeals to the general ordering of the reaction rates and depend on the nuclear physics input mostly via an atomic mass table (section \ref{Sec_ReacNetwork}).  We demonstrate that this approach reproduces the result of detailed reaction network by  \cite{lau_ea18} for initial iron composition very well and use it to study effects of nuclear physics uncertainties on composition and heat release in accreted neutron star crust (section \ref{Sec_results}).

\section{Simplified reaction network} \label{Sec_ReacNetwork}

Our model in general follows ref.\ \cite{Sato79,HZ90,HZ03,HZ08,Steiner12,lau_ea18,Fantina_ea18}. We consider nuclear reactions in accreted matter during compression by increasing pressure.
However, we do not 
enforce simplified assumption of one-component composition as in \cite{Sato79,HZ90,HZ90b,HZ03,HZ08,Fantina_ea18}, but we also do not track the evolution of nuclear reaction flow with time via detailed reaction network at is was done by \cite{lau_ea18}.
Instead, we apply the simplified stepwise reaction network, 
generally following the ideas suggested by  \cite{Steiner12}.

The algorithm of this stepwise network is following.
We start from the outer crust. The composition and pressure at this point are treated as an input parameter.
In this paper, we select initial pressure to be $P=6\times 10^{26}$~dynes\,cm$^{-3}$ and, following \cite{Steiner12}, start from pure $^{56}$Fe composition for simplicity. Our approach can be also applied for more complex initial composition, given by ashes produced in different types of bursts or during stable burning (see, e.g., \citealt{mdkse18} for a recent review), but we defer such calculations for future work.

With this as initial data, we start the stepwise increase of the pressure. 
For each value of pressure, we check all nuclei for energetically allowed  nuclear reactions 
from the list: (a) emission of neutrons,%
\footnote{Following \cite{lau_ea18}, we assume that the nucleus emits a maximal energetically allowed number of neutrons. See more detailed discussion below.}
 (b) neutron capture, (c) two-neutron capture, (d) electron capture
(e) pycnonuclear fusion.
The pycnonuclear reactions are treated as allowed, if their typical timescale is shorter than the compression rate, estimated as $P/(g\dot m)$, where $P$ is current value of pressure, $g$ is gravitational acceleration (in local Newtonian frame, assumed to be constant within the crust). Finally, $\dot m$ is a local accretion rate per unit element of the neutron star surface.
Following \cite{Steiner12}, we apply a simplified stepwise model for the kinetics of nuclear reactions:
at each step only a ``chunk'' (selected to be $\alpha = 0.001$ of total nuclei amount) of nuclei undergoes nuclear reaction. 
The same chunk is also applied to check reaction for energetic allowance by calculating the change of the Gibbs energy due to reaction, assuming that the pressure is the same in the parent and daughter state of the system.%
\footnote{As discussed by \cite{HZ90,Steiner12,Chamel_etal15_Drip}, it is useful to treat reactions in the crust of accreting neutron star as occurring at a fixed hydrostatic pressure of overlying material, so that the appropriate thermodynamic function to be minimized is the Gibbs energy.}
If at least one reaction is allowed, we reduce pressure to the threshold value to avoid additional energy release associated with the stepwise increase of the pressure. 
After that, we check if the reaction with the following chunk is allowed, while the pressure is kept fixed until none of the considered reactions can reduce the Gibbs energy.
Achieving this state, we make a new pressure increase step and repeat the procedure, described in this paragraph. The heating is calculated as a sum over energy release over reaction chunks. For each chunk, the energy release is assumed to be equal to the decrease of the Gibbs energy (for details see below).

A crucial point of our approach are the priority rules, which determine an order of stepwise reaction flow.%
\footnote{\cite{Steiner12} does not explain the priority rules explicitly, hence, a detailed comparison with the kinetic model of that paper is not possible. \label{Footnote_Steiner_model}
}
Our choice -- decreasing priority from (a) to (e)  is based on the  order-of-magnitude estimates of typical reaction timescales: the nuclear timescale ($\sim 10^{-21}$~s)  for the emission of an unbound neutron(s);
$\mathbf{10^{-18}-10^{-12}}$~s for exothermic neutron captures (e.g., \citealt{Goriely_ea08}); order of $10^{-5}-10^3$~s for electron captures (e.g.\ \citealt{Langanke_etal03}), and typically much longer for pycnonuclear reactions.
The stepwise reaction flow, controlled by the priority rules, mimics fast equilibration over the fastest reactions with subsequent equilibration over slower reactions, which should occur in a detailed reaction network. Similar idea of equilibration over fastest reactions is applied, e.g., in the SkyNet nuclear evolution code by \cite{SkyNet}, which automatically assumes nuclear statistical equilibrium over strong reactions, if corespondent reaction timescale becomes shorter than the timescale of the density changes and the temperature is high enough. Thanks to that, SkyNet does not need too small timesteps required to track strong reactions.
 
It should be noted, that
the preference between rates of (c) and (d) reactions is not so certain: within our approach the  reaction of type (d) occurs only, if one-neutron capture 
is not allowed energetically and requires thermal excitation.
However, if subsequent neutron capture is energetically favorable, it occurs very fast, leading to two-neutron capture as a net result. To account for this, we include reactions of this type (c) to our simplified kinetic approach. Its rate is determined primarily by the rate of the first neutron capture, which should  be calculated including neutron degeneracy and plasma effects (see \citealt{Shternin_ea12} for details). The resulting rate strongly depends on a neutron capture threshold and a neutron chemical potential. In realistic simulations it can be compared with electron capture rate, leading to reaction flow branching (see discussion in the section \ref{Sec_Lau}). However, here we assume it to be faster than the electron capture rate for the sake of simplicity and to avoid dependence of the model on other nuclear physics input except atomic mass tables.

Final set of priority rules is applied if reactions of the same type are allowed for several types of nuclei. Namely, we assume that the neutron emission takes place for the most abundant element; while
for reactions of types (b-d) we select the most energy-efficient one to go first.  
As for pycnonuclear reactions, we directly compare the reaction rates to determine the fastest one.

Several additional points should be specified, so that our results can be reproduced.
First, we neglect finite temperature effects for reactions of type (a-d). This means that only exothermal reactions are allowed. For pycnonuclear reactions we include temperature enhancement according to approach suggested in section III G by \cite{Yakovlev_ea06}; the required astrophysical factors are taken from \cite{Afanasjev_ea12}.
Following \cite{lau_ea18}, the temperature is assumed to be $T=5\times 10^8$~K in the whole crust.
Pycnonuclear reactions between nuclei of the same type are forbidden, if this type of nuclei exists in a fraction of just one chunk. This rule is applied to
reduce dependence of simplified reaction network on choice of chunk size $\alpha$ (the reaction rate for fusion of nuclei with fraction of one chunk is $\propto \alpha^2$, being vanishing at $\alpha\rightarrow 0$). Furthermore, it mimics realistic situation, when fast fusion is triggered by electron capture, leading to formation of low Z nuclei. Namely, if the pycnonuclear fusion of this newly formed low-Z nucleus with other nuclei in the same layer is fast enough, it will prevent the accumulation of such low-Z nuclei. As a result number density of such low-Z nuclei will be too small to got reasonable reaction rate between them. 
This scenario is a likely reason for domination of pycnonuclear fusion between different nuclei over fusion of two nuclei of the lowest $Z$ at some layers in detailed reaction network by simulation by \cite{lau_ea18} (for example, while describing reactions at $\mu_\mathrm{e}=37.1$~MeV for initial $^{56}$Fe composition the $^{28}$O$+^{40}$Mg fusion reaction is mentioned, but not  $^{28}$O$+^{28}$O).

Second, while calculating the threshold for electron capture we assume that it happens from the ground state of the parent nucleus to the ground state of the daughter nucleus. As discussed in \cite{HZ90} and following papers, this is not always true because the ground state - ground state transition can be forbidden and allowed electron capture should occur to the excited state.
These effects were included into models of accreting neutron stars' crust (e.g.\ by \citealt{Gupta_ea07, lau_ea18} and partly in \citealt{HZ90}), but it involves additional theoretical uncertainty associated with the prediction of the structure of the excited states.
Here we prefer to avoid it. Our primary goal is to check the dependence of models of the accreted neutron star's crust on the atomic mass tables.
Note, the uncertainty associated with the theory of excited states can be rather large. For example, the ground state -- ground state transition is forbidden for electron capture by $^{56}$Fe, thus increasing threshold electron chemical potential from $4.0$~MeV to $4.1$~MeV, if experimental data for excited states of $^{56}$Mn are applied (see, e.g., \citealt{HZ90}).%
\footnote{The quoted values are not equal to energy release in the beta decay of $^{56}$Mn in vacuum due to Coulomb energy $\mathcal E_\mathrm{C}$.}
 However, the use of a theoretical model for excited states of $^{56}$Mn defers electron capture to $\mu_\mathrm{e}=6.2$~MeV in the model applied by \cite{lau_ea18}.
Here and below we do not include electron mass to $\mu_\mathrm{e}$.

Third, the effects of electron captures to the excited states are not fully neglected, but included into the simplified reaction network on a qualitative basis.
As noted by \cite{Gupta_ea07}, the above-barrier electron captures often populate excited states of daughter nucleus, reducing the energy loss associated with neutrino emission.
To take this into account, we neglect neutrino energy losses (as shown in \cite{Gupta_ea07} by approximate reaction network, similar approach agreed pretty well with heating from their full network calculation) and assume that the daughter nucleus for electron capture reactions has excitation energy equal to the energy release in the electron capture. The latter assumption is required to avoid detailed consideration of the energy levels, as was stated above. We keep this excitation while returning to reactions of type (a) -- neutron emission.
Thanks to adopted from \cite{lau_ea18} assumption that maximal energetically allowed number of neutrons is emitted, it can lead to emission of additional neutrons and only after that the excitation energy of the final nucleus is supposed to be realized in form of heat.
Interestingly, that in some cases, the number of emitted neutrons is too large --- neutron captures become energetically favorable and proceed just after neutron emission. The same feature takes place in \cite{lau_ea18}.

It is worth to note, that if neutron abundance is not vanishing, the net result of this emission-capture process is typically the same as in assumption that neutrons are emitted subsequently (by one or pair) from the ground states of respective nuclei and excitation energy is released into the heat just after each chunk of reaction. 

Finally, in our simplified reaction network allowance of the reactions is controlled by the change of the Gibbs energy, thus it is crucial to specify the thermodynamics.
We neglect the finite temperature effects on thermodynamical functions and, following \cite{lau_ea18},  neglect the effects of the free neutrons on the atomic masses.
It allows to write down the energy density in the form
\begin{equation}
  \mathcal{E}=\mathcal{E}_\mathrm{i} + \mathcal{E}_\mathrm{e}+\mathcal{E}_\mathrm{n}+\mathcal{E}_\mathrm{C}.
  \label{E}
\end{equation}
Here
\begin{equation}
\mathcal{E}_\mathrm{i}=\sum_j M_j n_j \label{Ei}
\end{equation}
represents the energy density of nuclei,  $n_j$ is the number density of nuclei of type $j$.
We treat the nuclear masses,  $M_j$, as a nuclear physics input parameter for our model, and in the next section apply set of atomic mass tables.%
\footnote{Note, atomic mass tables traditionally report masses of the neutral atom,  i.e.\ values, quoted in these tables, includes the rest mass of electrons and electron binding energy. The latter was estimated as $1.433\times10^{-5} Z^{2.39}$~MeV (e.g., \citealt{FRDM12}) for all nuclei in all atomic mass tables applied in this work. Here $Z$ is nuclear charge.}
$\mathcal{E}_\mathrm{e}$ is an energy density of degenerate electrons (e.g., \citealt{ShapiroTeukolsky}). Exchange and polarisation effects are shown to be small by \cite{cf16_exchange} and are neglected here. 
For energy density of neutrons, $\mathcal{E}_\mathrm{n}$, we, following \cite{lau_ea18}, apply approximations from Table 5 of \cite{Sjoberg74}. The same parametrization was used in  \cite{MB77} compressible liquid drop model, adopted in models by \cite{HZ90,HZ90b,HZ03,HZ08}. Strictly speaking, choice of $\mathcal{E}_\mathrm{n}$ is additional nuclear physics input for our model and it should be chosen consistently with the applied atomic mass table. However, we don't expect that uncertainty associated with $\mathcal{E}_\mathrm{n}$ can affect the result crucially and apply \cite{Sjoberg74} model for all calculations here.

Finally, 
\begin{equation}
	\mathcal{E}_\mathrm{C}=-\sum_j \frac{9}{10}\frac{Z_j^{5/3}e^2}{a_\mathrm{e}} n_j, \label{EC}
\end{equation}
represents the Coulomb energy density of electron-ion system,%
\footnote{$\mathcal{E}_\mathrm{C}$ is often referred to as a \textit{lattice energy}, which is rather confusing because it is almost independent on the actual ordering on nuclei and, in particular, with rather good accuracy can be written in the same form for non-crystallized (liquid) state, which is actually the case for the outer crust.} 
calculated within ion-sphere model and linear mixing rule (see, e.g., \citealt{hpy07} and \citealt{Potekhin_ea09_linmixcor,kb15,cf16_mix} for more accurate calculations of $\mathcal{E}_\mathrm{C}$ in liquid and crystallized state).
Here $a_e=(4\pi n_\mathrm{e}/3)^{-1/3}$ is electron sphere radius. The electron number density $n_\mathrm{e}$ is given by quasi neutrality condition
$n_\mathrm{e}=\sum_j Z_j n_j$.

The pressure is given by derivative of the energy over volume at fixed composition, and for the energy density in form of  (\ref{E}) it can be written as
\begin{equation}
P=P_\mathrm{e}+P_\mathrm{n}+P_\mathrm{C},
\label{P}
\end{equation}
where $P_\mathrm{e}$ and  $P_\mathrm{n}$ is pressure of degenerate electrons and neutrons respectively, while the Coulomb correction is
$P_\mathrm{C}=\mathcal{E}_\mathrm{C}/3$.
Finally, the Gibbs energy density is $\mathcal{G}=\mathcal{E}+P$, while the Gibbs energy per nucleon is $g= \mathcal{G}/n_\mathrm{b}$. Here, the total baryon number density is $n_\mathrm{b}=n_\mathrm{n}+\sum_j A_j n_j$, where $n_\mathrm{n}$ is number density of free neutrons and $A_j$ -- number of nucleons in nucleus of type $j$.
 
By construction, our simplified kinetics model is almost independent of an accretion rate:
the only accretion-rate-dependent element is the allowance for pycnonuclear reaction by equivalence between their rate and compression timescale. However, thanks to very rapid (exponential) dependence of pycnonuclear reaction rates on the density, variation of the accretion rate for several orders of magnitude just slightly modifies the threshold pressure, where we allow pycnonuclear reaction to occur. For simplicity, we, following \cite{lau_ea18}, fix the accretion rate to $\dot m=8.8\times10^4$~g\,cm$^{-2}$\,s$^{-1}$ and $g=1.85\times 10^{14}$~g\,cm$^{-3}$ ($0.3$ of the Eddington accretion rate) 
in all calculations.

We stop our simulations at the same maximal pressure as in \cite{lau_ea18} -- $P=3.9\times 10^{30}$~dynes\,cm$^{-2}$.

\section{Results} \label{Sec_results}
\subsection{Comparison with detailed reaction network model by Lau et al. (2018)}\label{Sec_Lau}

As a test bench for our simplified reaction network, we choose the same initial conditions and mass model inputs as in the detailed calculations by \cite{lau_ea18}.
To reproduce them, we apply the atomic mass model as in that paper: atomic mass evaluation 2016 (AME16; \citealt{ame16})
 for experimentally known masses%
\footnote{The extrapolated mass values from this paper are ignored.}
and finite-range droplet macroscopic model (FRDM92, \citealt{FRDM92}) for most of the experimentally unknown masses.
For nuclei not included in this table we apply 31-parameter model by \cite*{DZ95,Anatomy_DZ_10}.%
\footnote{The table for AME16 and code for DZ31 model were downloaded from https://www-nds.iaea.org/amdc/\label{footnote_AME_DZ}. The table for FRDM92 model was downloaded from http://t2.lanl.gov/nis/molleretal/publications/ADNDT-59-1995-185-files.html}
Below we refer to the resulting mass table as `FRDM92+DZ31' model, omitting for brevity the name AME16, used in all calculations below.

Following \cite{lau_ea18} we do not mix masses from different tables (AME16, FRDM92, and DZ31) while calculating thresholds and Q-values. This means that we always apply a table, which includes masses for both parent and daughter nuclei for the considered reaction chunk, while calculating the change of the Gibbs energy after to reaction.
For example, for $A=56$ nuclei the element with lowest $Z$ reported at AME16 is $^{56}$Sc, which is formed as a result of electron capture by  $^{56}$Ti at $\mu_\mathrm{e}=15.8$~MeV (AME16 was applied to calculate this value, because both masses $^{56}$Ti and $^{56}$Sc are quoted there). Considering subsequent electron capture, we find that $^{56}$Ca is absent in AME16. So, not to mix the mass tables, we do not use the experimental mass for $^{56}$Sc, but calculate Q-value on the base of $^{56}$Sc and $^{56}$Ca masses reported in FRDM92.
The masses for nuclides in the most of the subsequent reactions (e.g., beta capture by $^{56}$Ca)  are reported in FRDM92, but not in AME16. Thus, masses of FRDM92 mass model is typically applied.
Below we refer to this approach for merging different mass tables as `unmixed'. In section \ref{Sec_Merg}, we also consider an alternative method (refered as `joint'), where different mass tables are simply combined into a joint mass table, which is applied in simplified reaction network as a universal mass table, without any special treatment of the boundaries of constituents.

\begin{figure}
	\includegraphics[width=\columnwidth]{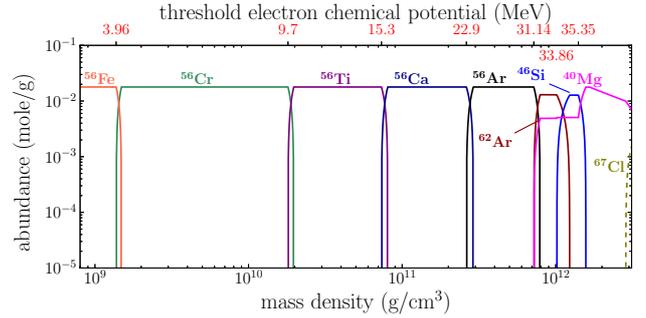}
	\caption{(color online) The composition of the accreted crust for FRDM92+DZ13 mass models as function of density in unmixed approach for merging mass tables (see text for details).}
	\label{Fig_CompareLau}
\end{figure}

Composition of the matter as a function of density for FRDM92+DZ31 mass model, obtained by our simplified reaction network approach, is shown in Fig.\ \ref{Fig_CompareLau}.
The electron chemical potentials, which correspond to the transitions, are marked on the top of the plot.

The figure is very similar to the results shown in Fig.\ 2 by \cite{lau_ea18}. In particular, the sequence of the most abundant elements is the same. The electron chemical potentials, which correspond to transitions, are a bit smaller in our simulation, because we assume ground state -- ground state electron captures, neglecting that such transition can be forbidden and allowed electron capture occurs to the excited state, requiring larger electron chemical potential.%
\footnote{\cite{lau_ea18} stated that they use a previous version of the atomic mass evaluation (AME12, \citealt{ame12}). For $A=56$ chain the update in AME16 is the inclusion of the mass of $^{56}$Sc, which was measured by \cite{ScMass}. Likely, this update was incorporated by \cite{lau_ea18}, because in opposite case the threshold for electron capture by $^{56}$Ti should be calculated within FRDM92 model, which predicts it to be $\mu_\mathrm{e}=16.7$~MeV (even for ground state-ground state transition), e.g. larger than the value quoted in table 1 of that work.}

Except for the absence of $^{54}$Ar, $^{58}$Ca, $^{61}$Ca, which appears in \cite{lau_ea18} at small ($\lesssim 10^{-4}$ mole\,g$^{-1}$) abundances at some narrow density regions, the main difference of out result is almost negligible production of $^{40}$Mg at $\rho\sim 10^{12}$~g\,cm$^{-3}$, associated with destroying of $^{62}$Ar by electron capture.

Let us discuss the reaction network in this density region in more details. The first electron capture by $^{62}$Ar triggers a cascade of electron captures and neutron emissions (superthreshold electron capture cascade, SEC, \citealt{gkm08}). As discussed in \cite{lau_ea18}, production of  $^{40}$Mg is associated with branching of the reaction flow on $^{42}$Si.
Namely, in our network at the beginning of $^{62}$Ar destruction, $^{42}$Si experiences electron capture, followed by neutron emission and finally arrives to $^{40}$Mg. However, the emission of neutrons increases neutron chemical potential. As a result, the channel with two-neutron capture on $^{42}$Si opens (one-neutron capture requires thermal activation with threshold $\sim 1.6$~MeV), and in our simplified network it is always preferable to electron captures. As a result, this branch drives nuclei to $^{46}$Si. In contrast to our approach, \cite{lau_ea18} calculated reactions rate and demonstrate, that the electron capture rate is not negligible.
It leads to more significant production of $^{40}$Mg, if compared to our results.

In principle, such feature can be incorporated into the simplified reaction network by more elaborated prescription for priority rules on the base of calculated reaction rates
(e.g., by dividing chunks into sub-chunks proportional to the reaction rate at the branching points of the reaction flow, where rates of two or more reactions are comparable). However, this will also introduce additional theoretical uncertainty to the model, associated with the calculation of reaction rates, which, in particular, requires modeling of the excited states structure for estimation of electron capture rates. As this work's aim was to check the dependence of accreted crust models on the uncertainties of the theoretical atomic mass tables, we prefer to keep our model as simple as possible. Additional nuclear physics input associated with detailed reaction rates is saved for the future work.

The formation of the first unbound neutrons is associated with electron capture by $^{56}$Ar at density $\rho\approx 8\times10^{11}$~g\,cm$^{-3}$. This electron capture triggers SEC, which drives to $^{40}$Mg. Part of the emitted neutrons is captured by $^{56}$Ar, which drives to the formation of $^{62}$Ar, in agreement with \cite{lau_ea18}. 

As in \cite{lau_ea18}, the pycnonuclear fusion occurs at density $\sim 2\times 10^{12}$~g\,cm$^{-3}$ in form of fusion-SEC cycle. Namely the compound nucleus, formed in pycnonuclear fusion, is rapidly destroyed by SEC sequence leading to $^{40}$Mg with the net result of conversion of one $^{40}$Mg nucleus into 40 neutrons.  However, the detailed kinetics of this process is a bit different. Namely, in \cite{lau_ea18} the main pycnonuclear reaction is $^{40}$Mg+$^{40}$Mg$\rightarrow ^{80}$Cr, accompanied by destruction of $^{40}$Mg to $^{25}$N by SEC  with subsequent  $^{25}$N+$^{40}$Mg$\rightarrow ^{65}$K pycnonuclear reaction. Both $^{80}$Cr and $^{65}$K are destroyed by SEC to initial $^{40}$Mg, with net result  $^{40}$Mg+$^{40}$Mg$\rightarrow ^{40}$Mg$+40$n for both reaction branches.
In our simulations, $^{40}$Mg  electron capture, which triggers SEC, occurs earlier, due to allowance for ground state -- ground state transitions.
As a result,  $^{40}$Mg+$^{40}$Mg reaction does not take place and the second  pycnonuclear branch is active, i.e. the main pycnonuclear reaction is fusion of lighter element, formed by SEC, with $^{40}$Mg. 
Generated heavy elements disintegrate into initial $^{40}$Mg by SEC, leading to the same net result of these reactions: the conversion of
$^{40}$Mg nuclei 
into $40$ neutrons, as both branches in \cite{lau_ea18}. 

Pycnonuclear-SEC cycle, along with compression, increases neutron number density and its chemical potential. Finally, at $3\times 10^{12}$~g\,cm$^{-3}$ it becomes high enough to open neutron capture on $^{61}$Cl. This isotope is formed in course of heavy elements disintegration by SEC, and neutron capture leads to the appearance of a substantial amount of stable $^{67}$Cl. Similar effect is predicted by \cite{lau_ea18}, when $^{46}$Si fraction builds up at large densities. It is important, that in our work all $^{46}$Si is destructed to $^{40}$Mg at densities, where no pycnonuclear reactions occur. The likely reason is that we neglect the energy levels of daughter nucleus, as stated above.   

Summarizing, we suppose that the general agreement of our simplified approach with detailed calculations is reasonable and can be used to check the sensitivity of the result to the mass models.

As in \cite{HZ90}, for FRDM92+DZ31 and all mass models applied in this paper,
the transitions in the crust, considered at a fixed pressure, can be accompanied by the density jump. Thus, further, for consistency, we report our results as function of pressure, in contrast to Fig.\ \ref{Fig_CompareLau}, where abundances of elements are shown as function of density, to make the comparison with Fig.\ 2 by
\cite{lau_ea18} more simple.

\subsection{Composition of the accreted neutron star for several atomic mass models}
\label{Sec_MassModels}

\begin{figure}
	\includegraphics[width=\columnwidth]{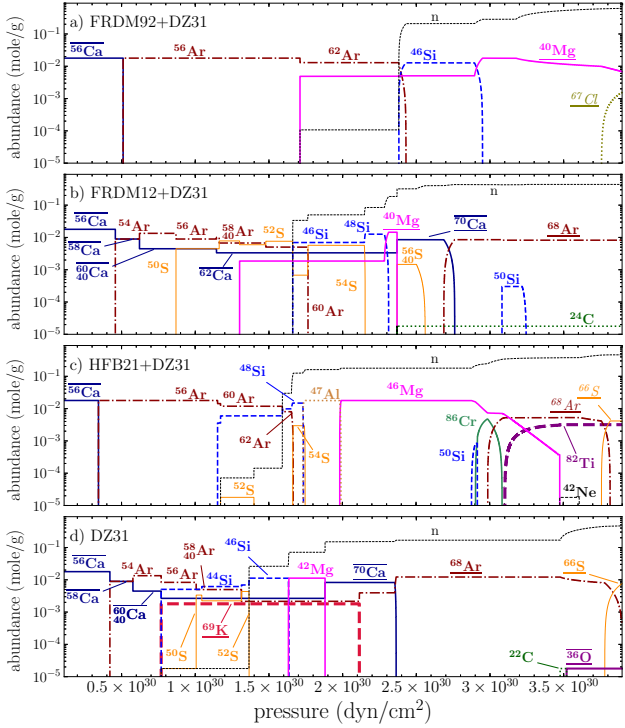}
	\caption{(color online) The composition of the accreted crust as function of pressure for  several atomic mass models in unmixed approach for merging the mass tables. Italic font indicates elements available only in DZ31 mass model. Overlining the isotopes means magic proton numbers while underlining -- magic neutron numbers. Also subshell gap N = 40 is specified (see the text for details). The shown abundances are the sum over all isotopes; the most abundant (everywhere dominant) isotope is indicated.}
	\label{Fig_unmixed}
\end{figure}

In this subsection, we apply the same approach, as in the previous one, fixing AME16 for experimental masses, but apply several mass tables, based on  different theoretical models, for nuclei, whose masses are not known from the experiment.
Namely, we use `FRDM12+DZ31', `HFB21+DZ31', and  `DZ31' models. In the first one updated finite-range droplet macroscopic model FRDM12
by \cite{FRDM12}
is used instead of FRDM92. In `HFB21+DZ31' FRDM92 is substituted by the mass table%
\footnote{Downloaded from http://www-astro.ulb.ac.be/bruslib/\\nucdata/hfb21-dat, \cite{bruslib}}
obtained within HFB approach, with the energy density functional selected to be BSk21 (Brussels-Montreal Skyrme)  by \cite{HFB21} to allow comparison with recent work by \cite{Fantina_ea18}.
Finally, in  
`DZ31' the DZ31 model is applied for all nuclides with experimentally unknown mass.
In the appendix we illustrate these models by mass charts. Namely, in Fig.\ \ref{Fig_MB}, to show the shell effects, they are compared with compressible liquid drop model by \citealt{MB77}. Fig.\ \ref{Fig_DZ} demonstrates difference of the applied models with DZ31 model.

The results of our simulations for these models are shown in Fig.\ \ref{Fig_unmixed}, panels (b-d). Panel (a) represents the same FRDM92+DZ31 model as in section \ref{Sec_Lau} as a function of the pressure, which is continuous driving parameter in our simulations. 

The difference between FRDM92 and FRDM12 consists in considerably improvement in the treatment of deformations and fewer approximations, which was required due to limited computer power available during work on \cite{FRDM92}. It should be also noted, that FRDM12 is fitted to the larger dataset -- AME12 experimental data by \cite{ame12}, while FRDM92 to the data available at that time (unpublished version of atomic mass evaluation, often refereed as AME1989). The update of the FRDM model significantly affects the composition of the accreted neutron star crust (compare panels a and b in Fig.\ \ref{Fig_unmixed}). Namely, the composition becomes at least two-component for significant part of the outer crust, since FRDM12 nuclear masses of Ca and Ar isotopes lead to the situation, when the capture of neutrons on $^{56}$Ca is more energetically favorable, than the same process on $^{54}$Ar. This branching drastically changes subsequent evolution of the composition in comparison with what follows from FRDM92. Improved consideration of deformation in \cite{FRDM12} generally diminishes the binding energy for nuclei with N = 40 subshell gap (see Fig.\ 6 in  \citealt{FRDM12}). However,  the relative changes in mass of isotopes leads to branching in outer crust with substantial amount of N = 40 elements. This feature was also caused by increased binding energy of $^{56}$S in FRDM12 compared to FRDM92, while N = 40 isotopes of Ar and Si became less bound. In addition, threshold electron chemical potential for $^{40}$Mg $e^-$--\,capture is 2.3 MeV lower than in FRDM92, this leads to SEC on $^{40}$Mg and pycnonuclear fusion of $^{34}$Ne\,+$^{34}$Ne and $^{34}$Ne\,+$^{24}$C. Only in deeper layers $P\approx 3.3\times10^{30}$~dynes\,cm$^{-2}$ the composition gets rid of impurities and becomes pure $^{68}$Ar. The appearance of $^{24}$C (or another light nuclei in further simulations) in the amount of one chunk is an artifact of our stepwise approach for pycnonuclear burning and does not affect the main results of the reaction network, being only a marker for the possibility that a small amount of unburned light elements can exist in the inner crust.  

Substitution of FRDM92 model to HFB21, based on different physical approach, modifies composition
substantially. It becomes much more complicated with larger number of components occurring after branching on $^{56}$Ar. Magic number of neutrons $N=28$ seems to be not enough strong to funnel  composition to $^{40}$Mg, as $N=28$ nucleus $^{42}$Si is less bound than $^{44}$Si. This leads to appearing of Si and S isotopes. S nuclides have more binding energy in HFB21 table in comparison with FRDM92. HFB21 table is limited to nuclei between neutron and proton drip lines, being significantly tighter than FRDM92 and FRDM12 tables (see Fig.\ \ref{Fig_DZ}). As a result, HFB21+DZ31 simulation is much stronger affected by data from DZ31 model and by treatment of the boundary between HFB21 and DZ31 model
(see section \ref{Sec_Merg}).
In particular, significant fraction of odd $A$ nucleus $^{47}$Al predicted to exist at $P\approx 1.8\times10^{30}$~dynes\,cm$^{-2}$  ($\rho\approx 10^{12}$~g\,cm$^{-3}$), however this nuclide is located in boundary of the applied HFB21 table, requiring thus usage of DZ31 model to calculate Q-value for electron capture in unmixed approach for merging of mass tables.
As expected, the reaction sequence and respective threshold chemical potential in the outer crust are in good agreement with the results by \cite{Fantina_ea18} for the same mass model. In contrast to FRDM92+DZ31 model,  there are no clear tendency to composition purification at pressure $P\sim3\times10^{30}$~dynes\,cm$^{-2}$, correspondent to the density $\sim 2\times 10^{12}$~g\,cm$^{-3}$ in this model. The reason is the stability (with respect to SEC) of high-Z nuclides lying at the boundary of HFB21 (see Fig.\ \ref{Fig_DZ}). These nuclides are formed as a result of $^{46}$Mg\,+$^{46}$Mg and $^{46}$Mg\,+$^{42}$Ne pycnonuclear burning reactions.

Results for DZ31 (panel d in figure \ref{Fig_unmixed}) confirm significant dependence of the composition of the accreted neutron star on the assumed mass model, however, they are rather similar to FRDM12+DZ31 model, showing significance of closure shell N = 50 (see also Fig.\ \ref{Fig_MB}). At pressure $P\sim (0.5-2.5)\times10^{30}$~dynes\,cm$^{-2}$ [$\rho\sim (2-15)\times10^{11}$~g\,cm$^{-3}$] the crust is predicted to be a complicated mixture and the most abundant isotopes are significantly not the same as in other mass models.
However, at larger density matter becomes almost pure, as in the case of FRDM92+DZ31, but dominated by $^{68}$Ar instead of $^{40}$Mg. Approaching to maximal pressure of our simulation $3.9\times10^{31}$~dynes\,cm$^{-2}$,  $^{68}$Ar gradually converts to $^{66}$S and composition becomes two-component consisting of two elements with N = 50. Local minimum at this closure shell with $Z\leqslant 20$ is demonstrated in Fig.\ \ref{Fig_MB}. It is noteworthy that DZ31 model overestimates significance of neutron magic numbers, including N = 50 (see top panel of Fig.\ \ref{Fig_DZ}), nevertheless composition naturally comes to N = 50 in all simulations.

\subsection{Merging the  mass tables}\label{Sec_Merg}

\begin{figure}
	\includegraphics[width=\columnwidth]{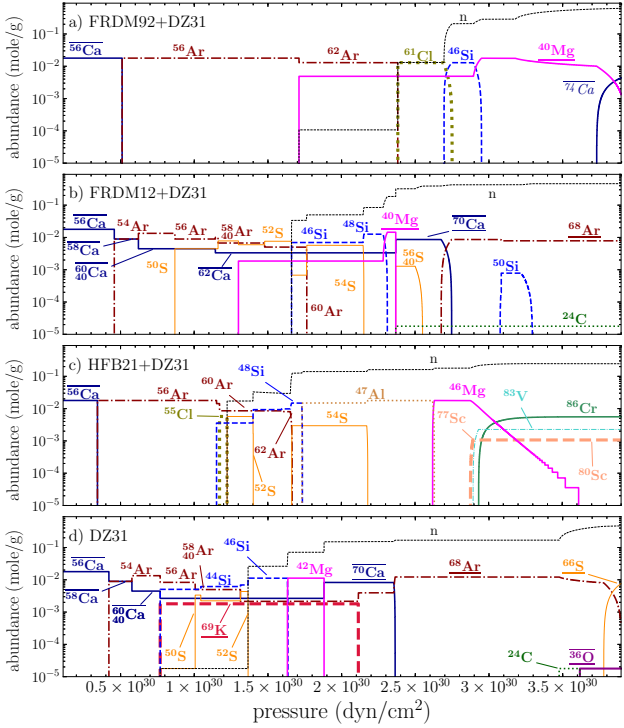}
	\caption{The same as in figure \ref{Fig_unmixed}, but for joint approach for merging the mass tables}
	\label{Fig_joint}
\end{figure}

The unmixed approach for merging the mass tables is expected to lead to more accurate results for the threshold of the nuclear reactions near boundaries of the mass tables, due to the absence of summation of the errors, which occurs, if different mass tables are applied for parent and daughter nuclei.
However, at least in realization applied in this work (section \ref{Sec_Lau}) it can have some artificial effects in thermodynamics. Namely, if reaction drives nucleus outside the previous model, we change the mass model to a new one  for the parent state of this nucleus (e.g., from AME16 to FRDM92 for example considered in the section \ref{Sec_Lau}), which leads to artificial change of the energy density of the nuclei system $\mathcal{E}_\mathrm{i}$ in parent state.

In principle, this effect can be avoided by shifting all masses in the new mass model to adjust the mass of the parent nucleus (e.g., by adding $2.8$~MeV to all masses of FRDM92 mass model to adjust the mass of $^{56}$Sc in AME16). However, it will damage the new mass model with regard to the mass of other experimentally known nuclei. Furthermore, if the reaction network will return to the experimentally known nucleus at some stage (e.g., due to neutron emission),
the experimental mass table should be also shifted, which can lead to the existence of nuclei with the same $Z$ and $A$ but different mass, depending on the path of their formation, which seems to be very unnatural.

Even worse, the unmixed approach can lead to the formation of closed loops of exothermal reaction flow, which works as \textit{perpetuum mobile} of the first kind.
For example, let us consider a full AME16 mass table, including extrapolated masses (we denote it as AME16\#) merged with FRDM92 mass table.
In this case, at neutron chemical potential $\mu_\mathrm{n}=1$~MeV we got the following cycle:
$^{41}$Al subsequently captures two neutrons and becomes $^{43}$Al, remaining within AME16\# model (total Q-value is $4.54$~MeV). 
The next neutron capture drives nucleus outside AME16\# data and within `unmixed' approach we consider it using FRDM92 mass models for both  $^{43}$Al
and  $^{44}$Al. The result is that the neutron capture is energetically allowed (Q-value $0.16$~MeV) and thus proceeds.
However, the resulting $^{44}$Al is not stable with respect of three-neutron emission (Q-value $0.74$~MeV), if it is considered within FRDM92 mass model for both  $^{44}$Al and $^{41}$Al isotopes, as it should be to avoid mixing of different mass models. As a result, we return to the initial  $^{41}$Al nucleus and close the cycle. The net energy release of this cycle is $5.44$~MeV per one $^{41}$Al nucleus going through it.
It should be noted, that simulations in Fig.\ \ref{Fig_unmixed} are free from such closed loops, but they appear naturally, if full AME16\# mass table with extrapolated data is applied. 

To avoid such peculiar behavior, we consider an alternative approach to merging the mass tables, where they are combined into one table, which is then directly applied to the simplified reaction network, without any special treatment of the boundaries of constituent mass models.
This approach can result in larger errors in the calculation of thresholds and Q-values for reactions that crosses the boundaries of the mass tables, but it is fully thermodynamically self-consistent, because it applies the same joint mass table throughout all stages of simulation.
Respective results are shown in Fig. \ref{Fig_joint}.

One can see that changing mass table merging approach also affects the composition. As noted in section \ref{Sec_MassModels}, the effect is the strongest for HFB21+DZ31 model, because the HFB21 mass table is rather narrow and the HFB21/DZ31 boundary is crossed often during simulation. In particular, appearance of Cl isotopes in HFB21+DZ31 as well as in FRDM92+DZ31 models is likely influenced by inconsistency of the respective mass tables at the boundary (see Fig.\ \ref{Fig_DZ}). The FRDM12+DZ31 and pure DZ31 mass models are less affected by the merging approach, because almost all reaction network at $P\gtrsim 10^{29}$~dynes\,cm$^{-2}$ apply nuclear masses from FRDM12 and DZ31 respectively, with few exceptions (for example, in DZ31 simulation light nuclei as $^{22}$C available in experimental data).

It is interesting, that for FRDM92+DZ31 model $^{74}$Ca instead of $^{67}$Cl is formed at at the end of simulation ($P\sim 3.9\times10^{30}$~dynes\,cm$^{-2}$) as a result of changing from unmixed to joint merging of the mass model. For HFB21+DZ31 model joint approach predicts rather complicated composition of high-Z elements at $P\sim 3.9\times10^{30}$~dynes\,cm$^{-2}$ located close to the boundary of HFB21 mass table and not rolling down to any known magic numbers.
It is likely associated with formation of local minima at N $\approx60$ at the borderline of HFB21 and DZ31 (see Fig.\ \ref{Fig_DZ}) as a result of joining of different mass tables.

\subsection{Profiles of the heat release and impurity parameter} \label{Sec_Heat_and_Imp}

\begin{figure}
	\includegraphics[width=\columnwidth]{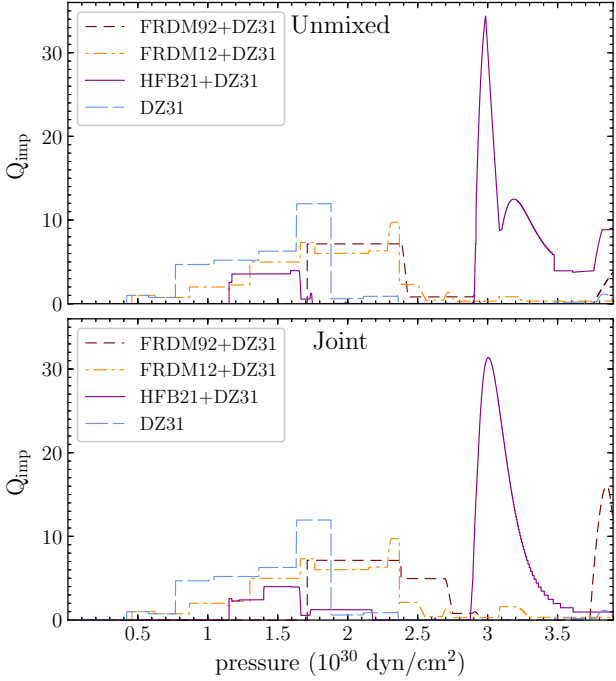}
	\caption{The profiles of the impurity parameter $Q_\mathrm{imp}$ for the models applied in this work.}
	\label{Fig_Qimp}
\end{figure}

Profiles of heat release and thermal conductivity are essential to model thermal evolution of the transiently accreting neutron star. Thermal conductivity is affected by the crust composition, which is typically parameterized via mean charge $\langle Z \rangle=\sum_j Y_j Z_j $ and the impurity parameter
\begin{equation}
 Q_\mathrm{imp}=\sum_j Y_j \left(Z_j-\langle Z \rangle\right)^2.
\end{equation}
Here $Y_j=n_j/\sum_j n_j$ -- number fraction of nuclei of type $j$.

The impurity parameter profiles are shown in Fig.\ \ref{Fig_Qimp}. For FRDM92+DZ31 model within the unmixed approach for merging the mass tables the impurity parameter profile agrees well with results by \cite{lau_ea18},%
\footnote{Small jumps of the $Q_\mathrm{imp}$ at the outer crust takes place at the transition regions, which are visible in Fig.\ 24 by \cite{lau_ea18}, but not shown in our plot, which represents the profile as function of pressure. The reason is that they do not occur in the structure of crust within our model (see, e.g., \citealt{cf16_mix} for discussion of transitions in case of ground state crust).}
 demonstrating increase at $P \sim 1.7\times 10^{30}$~dynes\,cm$^{-2}$. The same feature takes place for FRDM12+DZ31 and DZ31 mass models at lower pressures, corresponded to branching of the nuclear reaction chain. Peculiar behaviour demonstrates HFB21+DZ31 model, showing sharp increase up to values $\approx 35$ at $P \sim 3\times 10^{30}$~dynes\,cm$^{-2}$. This behaviour reflects reaction sequences for HFB21+DZ31 model discussed in section \ref{Sec_MassModels}.  For FRDM92+DZ31 model $Q_{\rm{imp}}$ have a peak near the end of the simulation, where increased neutron chemical potential stops SEC sequence for high $Z$ element formed by pycnonuclear fusion. It should be stressed, that the resulting nuclide is available only in DZ31 table and its stability depends on the approach of merging of the mass tables.

\begin{figure}
	\includegraphics[width=\columnwidth]{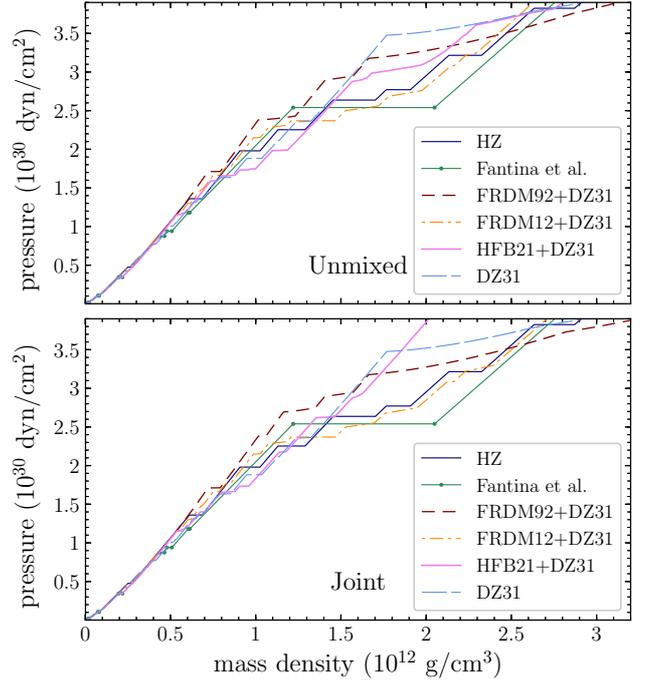}
	\caption{Equation of state for various models studied in the paper. Equation of state obtained by \protect\cite{HZ08} (denoted as HZ) and \protect\cite{Fantina_ea18} are shown as benchmarks.}
	\label{Fig_EOS}
\end{figure}

\begin{figure}
	\includegraphics[width=\columnwidth]{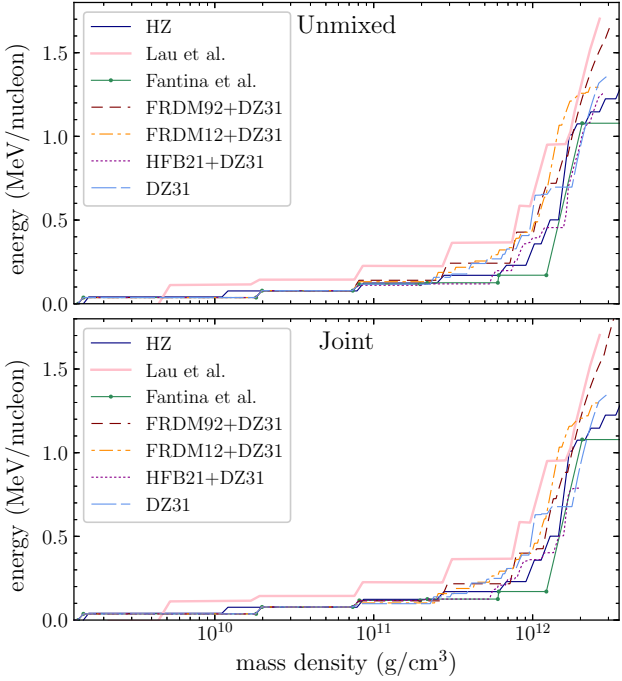}
	\caption{The profiles of accumulated heat release for the models applied in this work.}
	\label{Fig_heat}
\end{figure}

Another crucial parameter of the neutron star crust is equations of state, displayed in Fig.\ \ref{Fig_EOS} for various models. Showing almost similar behavior at low densities in the set of models, $P(\rho)$ curves substantially differ from each other near the end of the simulation, which corresponds to different reaction sequences and can be tracked in the profiles of integrated heat Fig. \ref{Fig_heat} (increase of energy liberated in reactions coincides with density jumps). 

The profile of accumulated heating (i.e. all heat released above a given layer) is drawn as a function of density to simplify comparison with the plots in the previous papers. Figure \ref{Fig_heat} includes all models considered in this paper along with results by \cite{lau_ea18}, \cite{HZ08} and \cite{Fantina_ea18} for initial $^{56}$Fe composition.
One can see that the profile is rather sensitive to the assumed mass model, but in general it resides between the results by \cite{lau_ea18} and \cite{Fantina_ea18}.
It should be noted, that \cite{lau_ea18} got additional heat at the first electron capture by $^{56}$Fe, because in their model it is delayed to $6.2$~MeV, following the theoretical model of the excited states of $^{56}$Mn applied in that work (see discussion in section \ref{Sec_ReacNetwork}). This delay leads to release of additional $\sim 2$~MeV at subsequent electron capture by $^{56}$Mn, with net result of additional $\sim 36$~keV/nucleon. A similar mechanism leads to additional heating for subsequent electron captures with the forbidden ground state-ground state transition. It seems to be the main reason for the difference between their results and our simulations for the same FRDM92+DZ31 mass model within the unmixed approach for merging the mass tables. The less power heating in \cite{Fantina_ea18} is not only caused by ground state -- ground state transition, but it is also due to considering thresholds for simultaneous electron capture and neutron emission (neutrons are not confined in the clusters after electron captures), preventing thus additional heat release from neutron emission after electron capture.

It is worth to point out that the approach applied for merging the mass models also affects the heat release and impurity parameter profile.

\section{Summary and conclusions}
In the paper, we have studied the effects of nuclear physics input on the models of accreted neutron star crust.
For this aim, we developed the approach of the simplified reaction network, based on the stepwise description of the kinetics of the nuclear reactions according to a set of priority rules (section \ref{Sec_ReacNetwork}).
The main nuclear physics input for our model is the atomic mass table, which should be supplemented by a model of the energy density of pure neutron matter $\mathcal E_\mathrm{n}$.
In this work, following \cite{lau_ea18}, the latter was assumed to be the same as in \cite{MB77}, i.e. parametrized by table 5 of \cite{Sjoberg74}.

In section \ref{Sec_Lau}, our approach reproduces sufficiently the results of recent calculations within the detailed reaction network by \cite{lau_ea18}, if the same atomic mass table is applied (AME16+FRDM92+DZ31), along with the same treatment of boundaries of these mass tables (unmixed).

In section \ref{Sec_MassModels} we demonstrate that the use of alternative mass models for nuclei with experimentally unknown masses (FRDM12+DZ31, HFB21+DZ31, DZ31) significantly affects the composition of the crust, as it follows from our simulation. The changes in the composition, associated with changes of mass models are more profound than the difference between our model and detailed reaction network.

In section \ref{Sec_Merg} we concentrate on the proper treatment of the boundaries of the mass models. We show that the unmixed approach, when we do not mix the mass tables while calculating thresholds and Q-values, can lead to closed cycles of nuclear reactions with non-vanishing energy release. As an alternative, we suggest a `joint' approach, which applies the 
joint mass table without any special treatment of boundaries of constituent mass tables.
The composition of the accreted neutron star crust shown to be affected by switching over from unmixed to joint approach.
However (see section \ref{Sec_Heat_and_Imp}), in spite of significant dependence of the crust composition on the mass model, the profile of the heat release is rather stable, confined between results by \cite{lau_ea18} and \cite{Fantina_ea18} for all considered models. The impurity parameter is affected stronger, but stays less than 35 for all models in consideration. It is worth to note that  the highest $Q_{\rm{imp}}$ values obtained in our simulations (a peak for  HFB21+DZ31 simulation and increase near the end of FRDM92+DZ31 calculation) can be associated with  effects of merging of different mass models, which can lead to formation of unnatural maximum of the binding energy not associated with any magic numbers (see Sec.\ \ref{Sec_Merg}). Consequently, to avoid such features it is very important to use one mass model for all experimentally unknown nuclei through the crust. Taking the DZ31 model as an example we can constrain impurity parameter to be less than 15 
and much lower for $P>2\times 10^{30}$~g\,cm$^{-3}$, there the matter tends to purification being funneled into the shell closures, confirming main conclusion of \cite{lau_ea18}. However, the sensitivity of this result to the initial composition should be checked in subsequent work; here, the initial composition was assumed to be pure $^{56}$Fe and all mixtures are formed as a result of nuclear reactions. The importance of the initial composition for the evolution of the impurity parameter in the crust was discussed by \cite{lau_ea18}.

In current version our approach neglects the neutron transfer reactions, which consists in quantum tunneling of neutron from one nucleus to another (\citealt{SA72_n_transf,Chugunov19_n_transf}). These reactions can be important for evolution of realistic ashes, furthermore, the rate of these reactions is very sensitive to the neutron separation energies, leading to an additional dependence of the reaction network on the assumed atomic mass table. We plan to consider these effect in subsequent work.

Our approach can be also applied to calculate the composition at larger densities with the use of the mass models including the effect of free neutrons on the nuclear masses (e.g., \citealt{MB77,DHM00}). For some preliminary results see \cite{SC19}.

\section*{Acknowledgements}
We are grateful to D.G.~Yakovlev, M.E.~Gusakov,  K.P.~Levenfish, and P.S.~Shternin for useful discussions and anonymous referee for insightful comments, which help us to improve the paper. 
Work is supported by Russian Science Foundation (grant 19-12-00133).


\appendix
\section{Masses in various models}

\begin{figure*}
	\begin{minipage}[l]{0.49\linewidth}
	\includegraphics[width=\linewidth]{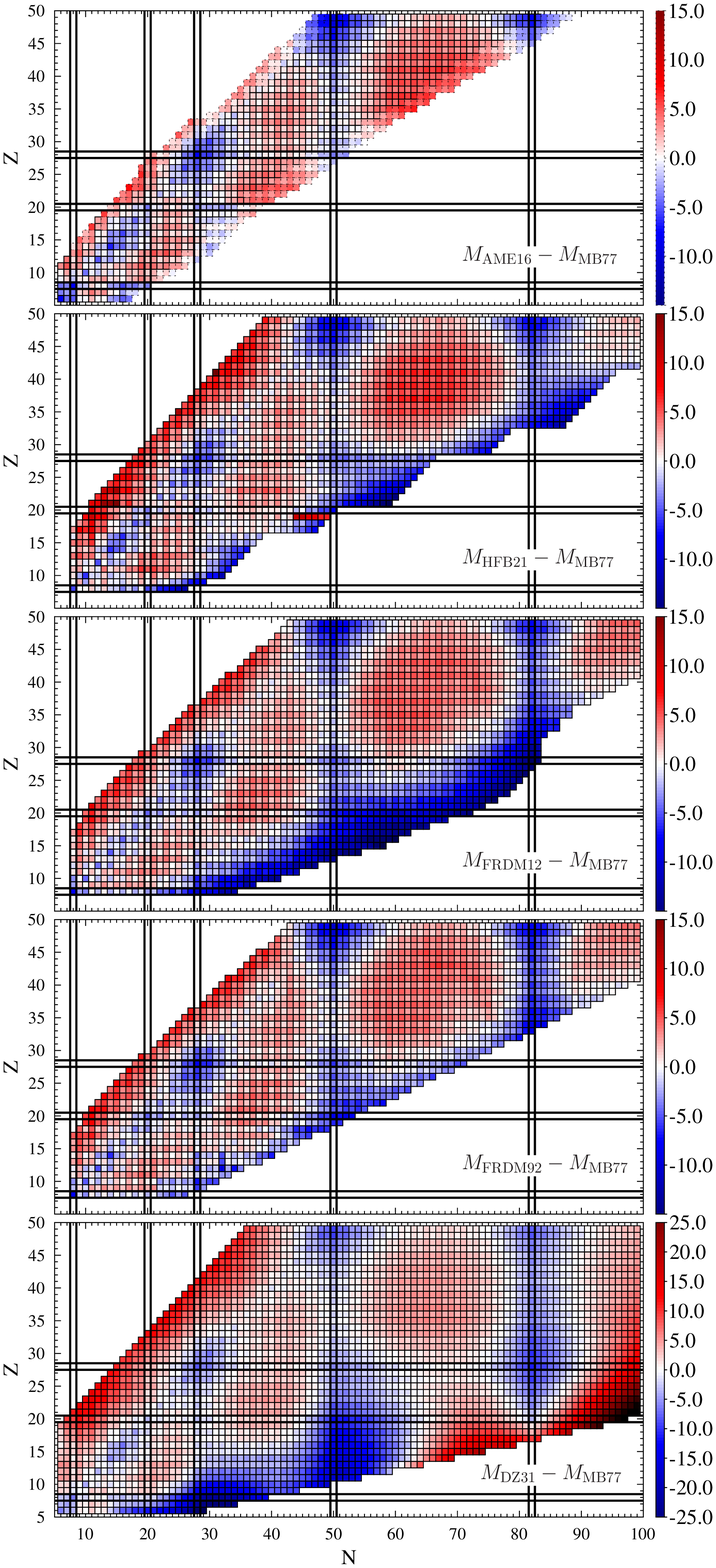}
	\caption{Mass differences [MeV] of applied mass tables with \protect\cite{MB77} model, which neglects shell effect. Magic numbers are shown by pair of thick solid lines and shell structure is well profound on the mass chart. Extrapolated data from AME16\# are demonstrated by dashed bounds at the top panel.}
	\label{Fig_MB}
	\end{minipage}
	\begin{minipage}[r]{0.48\linewidth}
	\vspace{-12.6pc}
	\includegraphics[width=\linewidth]{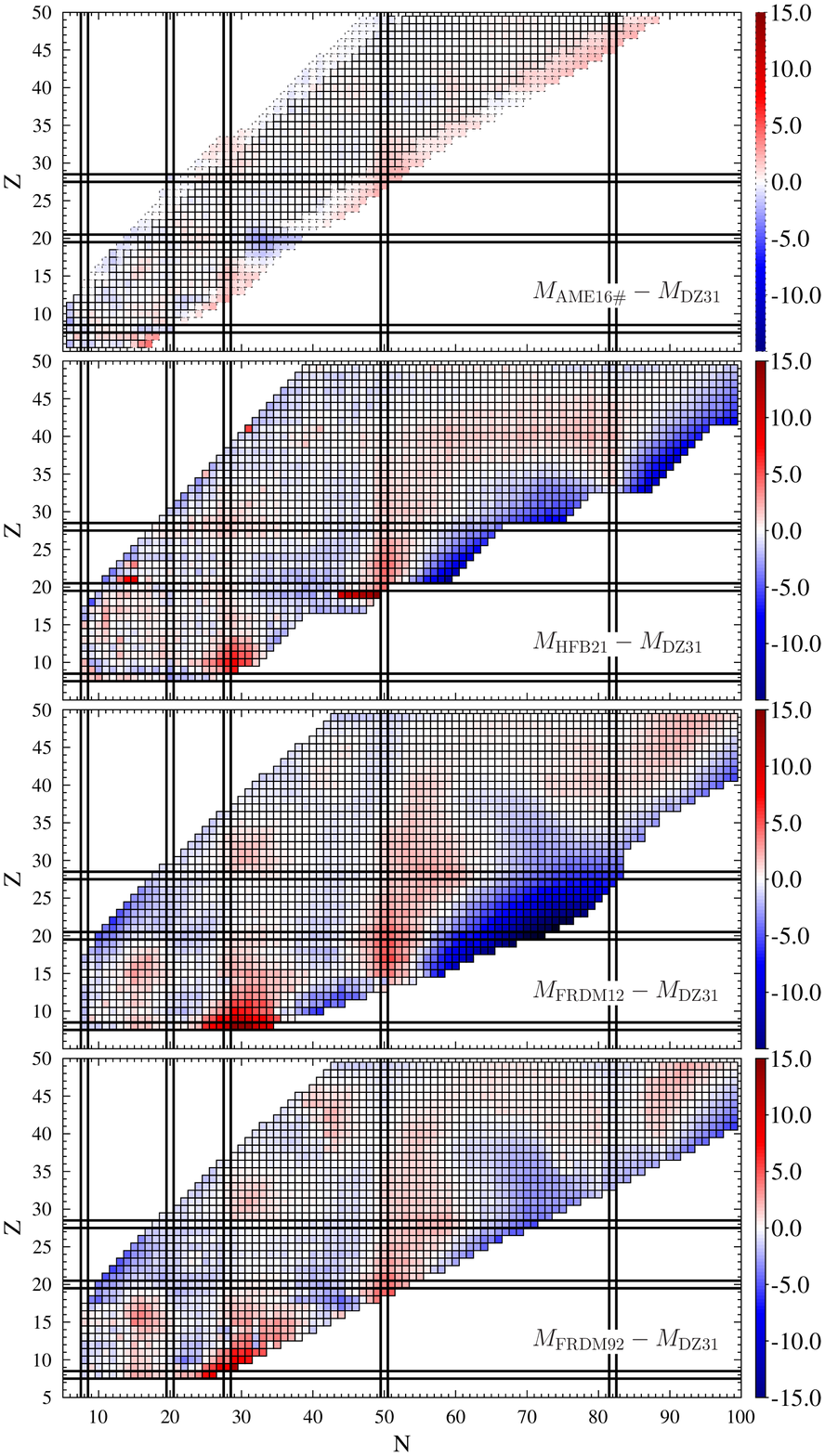}
	\caption{The same as Fig.\ \ref{Fig_MB}, but data from DZ31 model are subtracted from used mass tables}
	\label{Fig_DZ}
	\end{minipage}
\end{figure*}

\label{lastpage}

\end{document}